# Structural and electronic properties of rare earth chromites: A computational and experimental study

Jianhang Shi,[1,2] Gayanath W. Fernando,[3] Yanliu Dang,[2] Steven L. Suib,[2,4] and Menka Jain[2,3]

[1]Department of Materials Science and Engineering, University of Connecticut, Storrs, CT 06269, USA

[2]Institute of Materials Science, University of Connecticut, Storrs, CT 06269, USA

[3]Department of Physics, University of Connecticut, Storrs, CT 06269, USA

[4]Department of Chemistry, University of Connecticut, Storrs, CT 06269, USA

**Abstract:** In this work, the structural, optical, and electronic properties of rare-earth perovskites of the general formula $RCrO_3$, where R represents the rare-earth Gd, Tb, Dy, Ho, Er, and Tm, have been studied in detail. These compounds were synthesized through a facile citrate route. X-ray diffraction, Raman spectroscopy, and UV-Vis spectroscopy were used to reveal the structural evolutions in $RCrO_3$. The lattice parameter, $Cr^{3+}$-$O^{2-}$-$Cr^{3+}$ bond angle, and $CrO_6$ octahedral distortions were found to strongly depend on the ionic radii of the rare-earth element. First-principles calculations based on density-functional theory within the generalized gradient approximation (GGA) of Perdew- Burke- Ernzerhof (PBE) and strongly constrained-and-appropriately normed (SCAN) meta-GGA were also employed to calculate the structural and electronic properties of $RCrO_3$. The ground-state energy, lattice constants, electronic structure, and density of states (DOS) of $RCrO_3$ were calculated. These provide some insights into the electronic characteristics of the series of $RCrO_3$ compounds. The calculated values of lattice parameters and band gaps with Hubbard U correction (SCAN+U) agree well with values measured experimentally and show more accuracy in predicting the ground-state crystal structure and band structure compared to PBE+U approximation. The band gap of $RCrO_3$ is found to be independent of the ionic radii of the element R from both experiments and calculations.

## 1. Introduction

Oxides consisting of rare-earth element (R) with partially occupied 4*f* shell are crucial in modern technologies due to their various intriguing magnetic, luminescent, and electrochemical properties for potential applications as multifunctional materials.[1, 2] Investigations on oxides, such as rare-earth orthoferrites ($RFeO_3$),[3-5] manganites ($RMnO_3$),[6-8] nickelates ($RNiO_3$),[9-11] and orthochromites ($RCrO_3$),[12-14] have provided remarkable opportunities to enhance our understanding of the relationships between structural and physical properties.[15] Among the rare-earth oxides, the $RCrO_3$ with a distorted perovskite structure exhibiting unique interesting properties, such as spin reorientation, magnetization reversal, and large magnetocaloric effect, are currently attracting increasing research interest.[15] The $RCrO_3$ family of materials have shown to exhibit a G-type canted anti-ferromagnetism below *Néel* temperature, $T_N^{Cr}$, where the $Cr^{3+}$ spin orders.[16] The canting is presumed to be the source of weak ferromagnetism, which is the result of antisymmetric Dzyaloshinskii-Moriya (DM) interaction.[17] The transition temperatures ($T_N^{Cr}$) have been reported to shift to higher values with the increasing R-ion ionic radii, which could be attributed to the decreasing of lattice distortions and increasing $Cr^{3+}-O^{2-}-Cr^{3+}$ bond angles.[18] Recently, some of the rare-earth chromites were reported to be magnetoelectric multiferroics due to the coexistence of electric and magnetic orders, such as in $GdCrO_3$.[19] These properties allow $RCrO_3$ for possible practical applications such as spin-injection devices and nonvolatile magnetic random access memories. For periodic solids, first attention by the computational solid-state community goes towards predicting the fundamental band gap of materials due to its relevance in technological applications, such as opto-electronics and photovoltaics. $RCrO_3$ materials have also been studied as optically active photo catalyst due to the presence of slightly distorted $CrO_6$ octahedral complex.[17] Some materials of the $RCrO_3$ family have been reported to have wide band gap with a value ranging from 2.19 to 3.20 eV based on the results obtained from the UV–Visible spectroscopy.[17]

It is of great importance and value to utilize first-principles calculations to find out the lattice parameters of the stable crystal structure and its band structure in order to completely understand their electronic and magnetic properties.[20] Theoretically, first-principles computational techniques based on density functional theory (DFT) provide an extremely valuable tool for predicting structures and energetics of materials for both finite and periodic systems. Such techniques can assist in designing new compounds and aid in understanding the factors that can promote magnetoelectric coupling in multiferroics.[20] It should be noted that common current realization of DFT is through the Kohn-Sham approach, where the total energy of the system is determined variationally by a functional of the charge density.[21] However, the exact many-electron exchange-correlation part of this functional is unknown to this date.[21] Several approximations have been developed to allow for calculations at various levels of accuracy. The simplest approximation is the local-density approximation (LDA) functional, which uses a density functional that is based on the homogeneous electron gas whose exchange-correlation part has been estimated accurately by Monte-Carlo simulations.[21] A more sophisticated approximation to the exchange-correlation functional, generalized gradient approximation (GGA) in a standard form of Perdew-Burke-Ernzerhof (PBE), was developed after LDA, which incorporates its dependence not only on the electron density but also on its gradient.[21] More recently, a meta-GGA functional was developed to introduce a dependence on the second derivative of the density or Kohn-Sham orbital kinetic energy density in addition to the density and gradient.[22] The most successful meta-GGA to date is the Strongly Constrained and Appropriately Normed (SCAN) functional.[22] However, these approximations could sometimes dramatically underestimate the band gaps ($E_g$) for insulators due to the existence of a derivative discontinuity of the energy with respect to the number of electrons or the use of a local potential to represent exchange.[23-25] This inaccuracy could also be observed in the description of the electronic structure for strongly correlated systems.[26] In such cases, an efficient and commonly used tool to improve the accuracy of the DFT approximations is to incorporate a Hubbard-model-type correction (+U, correction to approximate DFT functionals such as, e.g., LDA, GGA, or meta-GGA), in which an empirical on-site potential (U) is added to the atomic pseudo- potential to account for localized $d$ and $f$ orbitals.[27] For many magnetoelectric multiferroic materials, such as $TbMnO_3$, $HoMnO_3$, the first-principles results have been shown to be very sensitive to the choice of the on-site potential.[28] However, spin-polarized versions of such density functionals have to be implemented when studying magnetic materials.

There are only few reports on the standard calculations based on DFT to understand the band structure of the $RCrO_3$. For example, Terkhi *et al.* reported a band gap of 2.15 eV of $GdCrO_3$ calculated by the modified Becke–Johnson (mBJ) exchange potential using the WIEN2K code,[29] which is lower than experimentally measured value of 3.15 eV.[17] For $DyCrO_3$, band gap was calculated to be 2.7 eV where for the exchange and correlation energy, the PBE functional under GGA was employed via Vienna *ab initio* simulation package (VASP), which is also lower than experimental value of 3.19 eV.[17, 30] This calculated band gap of 2.7 eV is attributed to the charge transfer gap between O $2p$ and Cr $3d$ states. In the work by Ong *et al.*, three optical gaps were identified: a charge transfer gap of 3.40 eV, a gap of 2.15 eV responsible for its green color, and an energy band gap of 1.40 eV between the occupied Cr $t_{2g}$ and unoccupied Cr $e_g$ orbitals.[31] However, the band gap of 2.15 eV is not revealed by the reflectivity measurement of pure $LaCrO_3$, which would have been a colorless material. This is in contradiction to the observed light green color. Although the aforementioned efforts have been made, the electronic structure of some $RCrO_3$ compounds remain to be understood from a theoretical point of view. Such theoretical work can lead to useful insights of the underlying physics of the evolution of their electronic structure and provide crucial information that can lead to the design of materials with enhanced physical properties, such as multiferroic properties.

In this article, we report the structural and electronic properties in the RCrO$_3$ family of materials (R = Gd, Tb, Dy, Ho, Er, and Tm) synthesized via the solution route. The optical band gap of ~ 3.3 eV is related to the transitions between the O 2$p$ valence band and the bottom of the conduction band. DFT based calculations are performed in the present work where the two exchange-correlation functionals were used within VASP: (PBE) GGA and (SCAN) meta-GGA functionals. The on-site Coulomb interaction correction was applied to both functionals that are now called as PBE+U and SCAN+U methods. To our knowledge, this is the first systematic study of computing their structural and electronic properties of the RCrO$_3$ family using DFT simulations within a collinear arrangement of R$^{3+}$ and Cr$^{3+}$ spins. The relationships among lattice modulation and electronic properties have also been investigated both experimentally and theoretically.

## 2. Experimental and computational details

RCrO$_3$ (R = Gd, Tb, Dy, Ho, Er, and Tm) polycrystalline samples were synthesized by citrate solution route. The high purity (> 99.99%) nitrate salts were obtained from Alfa Aesar. The metal salts were dissolved in water stoichiometrically and then mixed together with citric acid. The solution was continuously stirred, heated, and dried on a hot plate. The resultant powder was then grinded in mortar pestle and annealed at 900 °C for 2 h in oxygen atmosphere to obtain GdCrO$_3$ (GCO), TbCrO$_3$ (TbCO), DyCrO$_3$ (DCO), HoCrO$_3$ (HCO), ErCrO$_3$ (ECO), and TmCrO$_3$ (TmCO) bulk powder samples. The crystal structure of these powder samples was examined by X-ray diffraction (XRD, Bruker D2 Phaser diffractometer with Cu-Kα radiation) and by Raman spectroscopy (Renishaw System 2000 using 514 nm Ar-ion laser) techniques. The UV–Vis spectra of the samples were recorded using Shimadzu UV-2450 UV-Vis Spectrometer in a range of 200–800 nm using deuterium and halogens.

DFT-based spin-polarized electronic-structure calculations were carried out using projector-augmented wave method as implemented in VASP,[32, 33] with a kinetic energy cutoff of 520 eV and a total energy convergence threshold of 10$^{-6}$ eV.[30] In this work, the exchange correlation interaction is treated within the GGA using the PBE functional and within the meta-GGA using the SCAN functional, both with on-site Coulomb interactions (PBE+U, SCAN+U) for a better treatment of $3d$ (Cr$^{3+}$) and $4f$ (R$^{3+}$) electrons.[34] The structural properties and band structure of RCrO$_3$ are investigated here using PBE+U and SCAN+U and the two results are compared. The on-site Coulomb interaction presented in $3d$ states of the transitions metal-ion is corrected by the DFT+U (U is the Hubbard energy) method and we set U = 3 eV for Cr $3d$ states. The Hubbard U values for Gd, Tb, Dy, Ho, Er, and Tm are taken as 4.6, 5.0, 5.0, 4.9, 4.2, and 4.8 eV, respectively, according to the work by Topsakal *et al*.[35] We assume a 3+ oxidation state with $4f$ electrons of rare-earth frozen in the ionic core for PBE+U. For the SCAN+U, these $4f$ electrons are included as valence electrons and solved explicitly with the rare-earth atoms allowed to order magnetically (G-type antiferromagnetic).[20] The spin–orbit coupling and non-collinear magnetic states are not considered in these calculations. A $10 \times 10 \times 7$ Γ-centered k-point mesh was used to sample the Brillouin zone corresponding to the 20-atom orthorhombic cell. The structures are fully relaxed until the forces acting on the atoms were smaller than 0.005 eV/Å.

## 3. Results and discussion

### 3.1 Spin configuration

Three types of antiferromagnetic spin structures are possible for $R^{3+}/Cr^{3+}$ spins: G-AFM, C-AFM, and A-AFM.[36] Following the experimental evidence that the G-type spin structure is observed for both $R^{3+}$ and $Cr^{3+}$ sublattices from neutron elastic scattering measurements, in this work the G-AFM spin arrangement is considered for both $Cr^{3+}$ and $R^{3+}$ moments to save computation time.[36] Two typical spin configurations are proposed in the simulations: (i) $R^{3+}$ cations were treated as non-magnetic with the 4$f$ electrons frozen at the ionic core in simulations with PBE+U, (ii) $R^{3+}$ cations were allowed to order magnetically with the 4$f$ electrons treated as valence in simulations with SCAN+U. The two spin configurations for $Cr^{3+}$ and $R^{3+}$ sublattice were initialized in G-AFM order in a 20-atom unit cell as shown in Fig. 1. A full structural relaxation with the $R^{3+}$ and $Cr^{3+}$ magnetic moments initialized was conducted within a self-consistent field calculation of the electronic structure. The resulting ground-state configuration is analyzed focusing on the ground-state band structure, magnetic moments, and structural parameters.

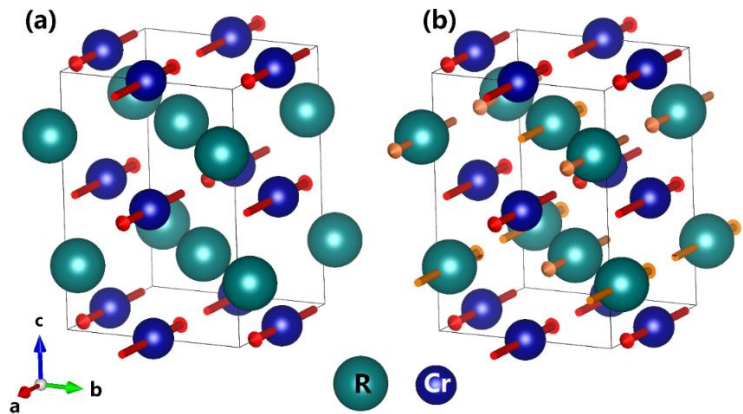

**Figure 1.** Spin configurations of R and Cr ions in RCrO$_3$ with: (a) $R^{3+}$ treated as non-magnetic ion with 4$f$ electrons frozen at the ionic core for the PBE+U and (b) $R^{3+}$ treated as magnetic ion with 4$f$ electrons solved explicitly for the SCAN+U.

Tables 1 and 2 summarize the ground state magnetic moments of ions in a representative material GdCrO$_3$ in RCrO$_3$, which was optimized by the SCAN+U and PBE+U simulations, respectively. In these tables, the contributions to the total (spin) magnetic moments of various electrons are listed, such as moments from $s$, $p$, $d$, and $f$ electrons. It is observed that the major contribution to total magnetic moments is from electrons of $d$ and $f$ shells for $Cr^{3+}$ and $Gd^{3+}$, respectively. The calculations predict that both $Gd^{3+}$ and $Cr^{3+}$ sublattice moments align antiferromagnetically in a collinear fashion. A zero net total moment of $GdCrO_3$ is observed in the present SCAN+U and PBE+U simulations. Similar results (negligible net total moments of $RCrO_3$) were reported in other rare-earth chromties in SCAN+U simulations, indicating the predominant antiferromagnetic spin configuration in $RCrO_3$ ground state.

**Table 1.** Details of the optimized magnetic moments in $GdCrO_3$ calculated by SCAN+U. The atomic positions of listed atoms are as follows: Gd (1): (0.982, 0.064, 0.250), Gd (2): (0.018, 0.936, 0.750), Gd (3): (0.482, 0.436, 0.750), Gd (4): (0.518, 0.564, 0.250), Cr (1): (0.500, 0, 0), Cr (2): (0, 0.500, 0), Cr (3): (0.500, 1.000, 0.500), Cr (4): (1.000, 0.500, 0.500).

| Atom | $s$ ($\mu_B$) | $p$ ($\mu_B$) | $d$ ($\mu_B$) | $f$ ($\mu_B$) | Total ($\mu_B$) |
|---|---|---|---|---|---|
| Gd (1) | -0.01 | -0.03 | 0.10 | 7.01 | 7.08 |
| Gd (2) | 0.01 | 0.03 | -0.10 | -7.01 | -7.08 |
| Gd (3) | -0.01 | -0.03 | 0.10 | 7.01 | 7.08 |
| Gd (4) | 0.01 | 0.03 | -0.10 | -7.01 | -7.08 |
| Cr (1) | -0.02 | -0.03 | -2.85 | 0.00 | -2.90 |
| Cr (2) | 0.02 | 0.03 | 2.85 | 0.00 | 2.90 |
| Cr (3) | 0.02 | 0.03 | 2.85 | 0.00 | 2.90 |
| Cr (4) | -0.02 | -0.03 | -2.85 | 0.00 | -2.90 |
| Total | 0.00 | 0.00 | 0.00 | 0.00 | 0.00 |

**Table 2.** Details of the optimized magnetic moments in $GdCrO_3$ calculated by PBE+U. The atomic positions of listed atoms are as follows: Gd (1): (0.982, 0.065, 0.250), Gd (2): (0.018, 0.935, 0.750), Gd (3): (0.482, 0.435, 0.750), Gd (4): (0.518, 0.565, 0.250), Cr (1): (0.500, 0, 0), Cr (2): (0, 0.500, 0), Cr (3): (0.500, 1.000, 0.500), Cr (4): (1.000, 0.500, 0.500).

| Atom | $s$ ($\mu_B$) | $p$ ($\mu_B$) | $d$ ($\mu_B$) | Total ($\mu_B$) |
|---|---|---|---|---|
| Gd (1) | -0.00 | -0.00 | 0.00 | 0.00 |
| Gd (2) | 0.00 | 0.00 | -0.00 | -0.00 |
| Gd (3) | -0.00 | -0.00 | 0.00 | 0.00 |
| Gd (4) | 0.00 | 0.00 | -0.00 | -0.00 |
| Cr (1) | -0.02 | -0.03 | -2.87 | -2.93 |
| Cr (2) | 0.02 | 0.03 | 2.87 | 2.93 |
| Cr (3) | 0.02 | 0.03 | 2.87 | 2.93 |
| Cr (4) | -0.02 | -0.03 | -2.87 | -2.93 |
| Total | 0.00 | 0.00 | 0.00 | 0.00 |

### 3.2 Structural properties

The room temperature XRD patterns of the RCrO$_3$ (R = Gd, Tb, Dy, Ho, Er, and Tm) polycrystalline samples are presented in Fig. 2, with corresponding Miller indices (hkl) of the characteristic peak.[15, 37] All the peaks can be successfully indexed in the measured 2-Theta range indicating that the samples are of single-phase with the space group *Pbnm*. The 2-Theta (2θ) positions of several representative peaks were summarized in Table 3. In general, the 2-Theta positions of most crystal planes shift to higher 2-Theta angles as the rare-earth atom changes from Gd to Tm (increasing atomic number and decreasing ionic radius). This indicates systematic structural variation, such as the decreasing of interplanar spacing, in the RCrO$_3$ series as a function of rare-earth size.

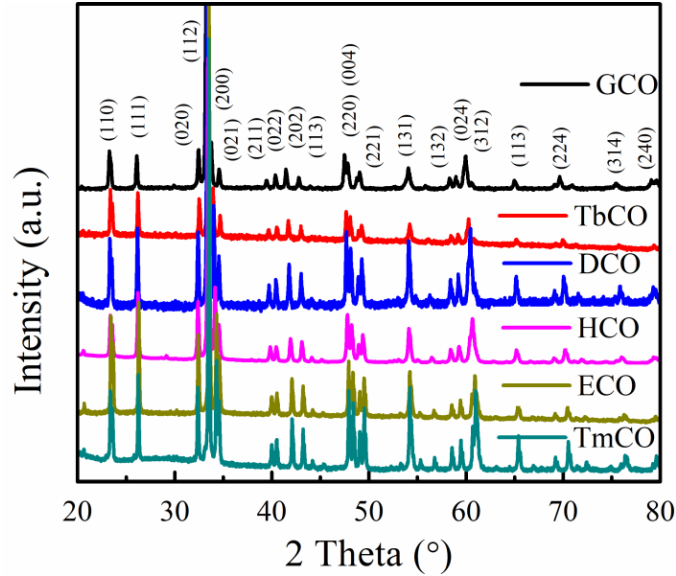

**Figure 2.** X-ray diffraction pattern of bulk rare-earth chromites (RCO) recorded under room temperature. The XRD peaks are indexed with Miller indices (hkl).

**Table 3.** Experimentally obtained 2-theta positions of the (111), (112), (202), (220), and (312) planes in RCrO$_3$ compounds.

| Sample | (111)° | (112)° | (202)° | (220)° | (312)° |
|---|---|---|---|---|---|
| GCO | 26.091 | 33.232 | 41.446 | 47.495 | 59.977 |
| TbCO | 26.204 | 33.373 | 41.673 | 47.653 | 60.278 |
| DCO | 26.153 | 33.363 | 41.784 | 47.673 | 60.460 |
| HCO | 26.220 | 33.428 | 41.931 | 47.775 | 60.674 |
| ECO | 26.294 | 33.525 | 42.076 | 47.915 | 60.904 |
| TmCO | 26.254 | 33.505 | 42.086 | 47.935 | 61.015 |

The experimentally obtained XRD scans were fitted for all samples by using Rietveld refinements via FullProf Suite software assuming an orthorhombically distorted perovskite structure (space group *Pbnm*) and four formula units per unit cell. Useful structural information was extracted from the refinements and the obtained lattice parameters, bond angles, and unit cell volumes are summarized in the Table 4. The lattice parameters (*a* and *c*) and unit cell volume were found to decrease with the decreasing size (ionic radii) of the rare-earth ion (as we go from Gd, Tb, Dy, Ho, Er, to Tm). The largest and smallest values of bond angles are observed in GCO and TmCO, respectively. The smaller and smaller values of the bond angles Cr-O$_1$-Cr (out of plane), Cr-O$_2$-Cr (in plane) from an ideal 180° of ABO$_3$ cubic perovskites structure corresponds to the tilting of CrO$_6$ octahedral and hence more and more distortion in the structure.[38] The out-of-plane (Cr-O$_1$-Cr) and in-plane (Cr-O$_2$-Cr) bond angles were found to decrease with decreasing atomic number of rare-earths, which is consistent with the decreasing size of rare-earth cations.[15]

**Table 4.** Ionic radii (IR) of rare earth ions and structural parameters of various RCrO$_3$ (RCO) compounds obtained by Rietveld refinement of the experimentally obtained XRD patterns: lattice parameters: *a*, *b*, *c*; unit cell volume: V; out-of-plane bond angle Cr-O$_1$-Cr, and in-plane bond angle Cr-O$_2$-Cr).

| Sample | IR(Å) | *a* (Å) | *b* (Å) | *c* (Å) | V (Å$^3$) | Cr-O$_1$-Cr (°) | Cr-O$_2$-Cr (°) |
|---|---|---|---|---|---|---|---|
| GCO | 1.107 | 5.319 | 5.527 | 7.614 | 223.839 | 151.4 | 149.7 |
| TbCO | 1.095 | 5.298 | 5.523 | 7.584 | 221.870 | 146.6 | 147.4 |
| DCO | 1.083 | 5.270 | 5.526 | 7.562 | 220.193 | 146.5 | 150.5 |
| HCO | 1.072 | 5.252 | 5.529 | 7.550 | 219.202 | 145.9 | 148.4 |
| ECO | 1.062 | 5.229 | 5.521 | 7.526 | 217.275 | 144.8 | 144.6 |
| TmCO | 1.052 | 5.213 | 5.512 | 7.508 | 215.737 | 147.0 | 143.6 |

The relaxed lattice parameters were also obtained from the DFT calculations. The calculated values of lattice parameters in Table 5 were obtained by using the initial spin configuration presented in Fig. 1(a) and PBE+U simulation, while in Table 6 those were calculated by SCAN+U and spin configuration in Fig. 1(b). Fig. 3 displays the experimental structural properties together with the computed results from Table 5 and Table 6. In general, the computed results from both PBE+U and SCAN+U predict a trend that is consistent with experimental lattice constants in this work.

Both the experimental and DFT-calculations show that the lattice parameters and unit cell volume decrease with the decreasing ionic radii of the rare-earth. The out-of-plane Cr-O$_1$-Cr angle is found to be decreasing with reducing ionic radii of rare-earth using both PBE+U and SCAN+U simulations. It is evident from Fig. 3 that the SCAN+U predicted lattice parameters are in better agreement with the experimental results compared to the PBE+U simulation. This indicates that meta-GGA functional facilitates comparatively better calculation of the structural properties of RCrO$_3$ perovskites. This observation agrees with previous study that the recently developed SCAN meta-GGA has shown to be superior to the local-density approximations/PBE GGA for predicting the geometries and energies of diversely bonded materials (including metallic, ionic, hydrogen, covalent, and van der Waals bonds).[39]

**Table 5.** Structural parameters obtained from the DFT (PBE+U) simulations.

| Sample | *a* (Å) | *b* (Å) | *c* (Å) | V (Å$^3$) | Cr-O$_1$-Cr (°) | Cr-O$_2$-Cr (°) |
|---|---|---|---|---|---|---|
| GCO | 5.340 | 5.614 | 7.679 | 230.246 | 145.60 | 146.66 |
| TbCO | 5.317 | 5.609 | 7.657 | 228.362 | 144.49 | 145.83 |
| DCO | 5.297 | 5.600 | 7.637 | 226.541 | 143.45 | 145.10 |
| HCO | 5.277 | 5.592 | 7.619 | 224.809 | 142.48 | 144.43 |
| ECO | 5.258 | 5.583 | 7.602 | 223.177 | 141.66 | 143.84 |
| TmCO | 5.237 | 5.569 | 7.582 | 221.147 | 140.59 | 143.12 |

**Table 6.** Structural parameters obtained from the DFT (SCAN+U) simulations.

| Sample | a (Å) | b (Å) | c (Å) | V (Å$^3$) | Cr-O$_1$-Cr (°) | Cr-O$_2$-Cr (°) |
|---|---|---|---|---|---|---|
| GCO | 5.320 | 5.577 | 7.637 | 226.565 | 146.42 | 147.26 |
| TbCO | 5.284 | 5.555 | 7.592 | 222.857 | 144.24 | 145.69 |
| DCO | 5.262 | 5.525 | 7.570 | 220.074 | 142.92 | 144.44 |
| HCO | 5.224 | 5.517 | 7.538 | 217.223 | 141.56 | 143.47 |
| ECO | 5.241 | 5.535 | 7.548 | 218.939 | 142.08 | 143.95 |
| TmCO | 5.199 | 5.506 | 7.514 | 215.081 | 140.27 | 142.61 |

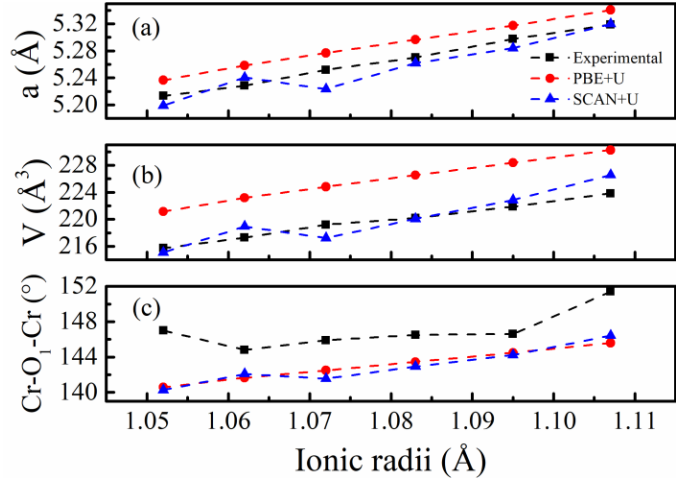

**Figure 3.** Comparison of (a) lattice parameter *a*, (b) unit cell volume (V) and (c) out-of-plane bond angle Cr-O$_1$-Cr obtained from DFT and Rietveld refinement of the experimental XRD data.

RCrO$_3$ with *Pbnm* space group contains four formula units per Bravais unit cell. Among 60 irreducible representation for vibration modes of atoms at four different Wyckoff sites, only 24 phonon modes (7A$_g$ + 7B$_{1g}$ + 5B$_{2g}$ + 5B$_{3g}$) are Raman-active.[40] Several characteristic Raman modes out of those 24 were observed in the Raman spectra of present samples as shown in Fig. 4. In these, several modes merged to form a broad peak for some RCrO$_3$ samples.[41] The assignment of phonon modes to the chromite spectra were carried out according to the work by Weber *et al.* and Camara *et al.*[41, 42] The Raman modes A$_g$ (~140cm$^{-1}$) and B$_{2g}$ (~160 cm$^{-1}$) are induced by displacements of the A-site ions since the heaviest atom of the structure is anticipated to vibrate at the low wave number region.[41] Bands in mid-spectral region (200-400 cm$^{-1}$), such as A$_g$ (~260 cm$^{-1}$), A$_g$ (~330 cm$^{-1}$), and A$_g$ (~400 cm$^{-1}$), are very sensitive to the changes in the orthorhombic distortion. Two modes A$_g$(3) and A$_g$(5) have been identified as octahedral rotation soft modes, as Raman shifts scale linearly with the tilt angle of the CrO$_6$ octahedra.[41] Both A$_g$(3) and A$_g$(5) are found to increase with the decreasing of the ionic radii of rare-earth, indicating a larger tilt angle for RCrO$_3$ with smaller ionic radii. This observation agrees with the results predicted by the DFT-predicted bond angle in Fig. 3(c). The shift of some Raman modes for all samples are summarized in Table 7. Some representative modes are plotted in Fig. 5 to show the trends that the Raman modes shifted to lower positions with the increasing of ionic radii.

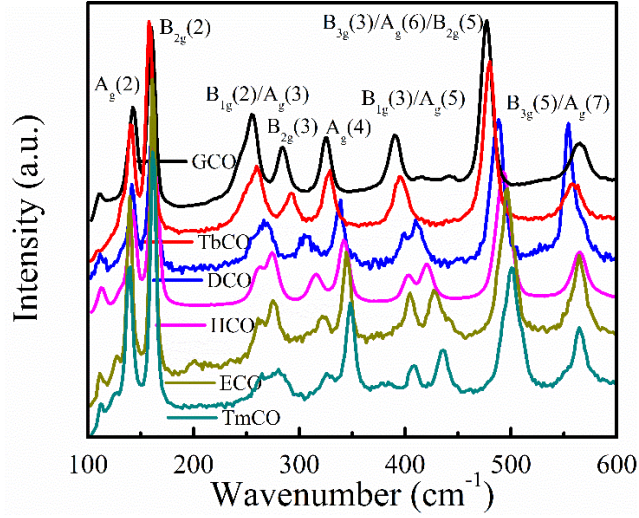

**Figure 4.** Room temperature Raman spectra of rare-earth chromites.

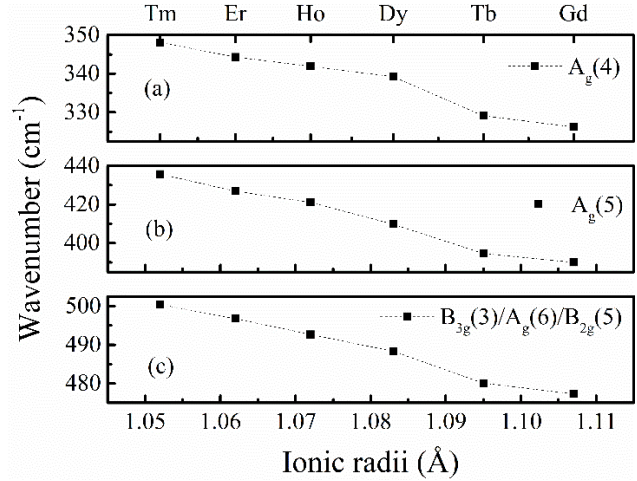

**Figure 5.** Evolution of the room temperature Raman modes' position with ionic radii of $R^{3+}$ in $RCrO_3$.

**Table 7.** Raman shifts of different modes.

| Sample | $B_{2g}(2)$ (cm$^{-1}$) | $A_g(3)$ (cm$^{-1}$) | $A_g(4)$ (cm$^{-1}$) | $A_g(5)$ (cm$^{-1}$) | $B_{3g}(3)/A_g(6)/B_{2g}(5)$ (cm$^{-1}$) |
|---|---|---|---|---|---|
| GCO | 160.243 | 254.964 | 326.219 | 390.039 | 477.263 |
| TbCO | 157.887 | 259.610 | 329.099 | 394.616 | 480.094 |
| DCO | 162.914 | 266.324 | 339.219 | 409.809 | 488.344 |
| HCO | 162.137 | 274.266 | 341.914 | 421.049 | 492.693 |
| ECO | 161.086 | 274.968 | 344.338 | 426.876 | 496.775 |
| TmCO | 161.354 | 280.455 | 348.055 | 435.690 | 500.429 |

### 3.3 Optic and Electronic properties

UV–Vis diffuse reflectance spectroscopy was used to probe the electronic behaviors present in the $RCrO_3$, such as electronic transitions of the different orbitals of a solid.[43] The reflectance spectra can be seen from Fig. 6(a). To determine the bandgap, the measured reflectance (R) needs to be converted to its corresponding absorption ($F_R$) according to the Kubelka–Munk function:[43]

$$F_R = \frac{(1-R)^2}{2R} \tag{1}$$

For the calculation of bandgap, the Tauc's equation is employed as below:[44]

$$\alpha h\nu = B(h\nu - E_g)^n \tag{2}$$

where $\alpha$ is the optical absorption coefficient, $h$ is the Planck constant, $\nu$ is the photon's frequency, B is a characteristic parameter, $E_g$ is the energy bandgap, n is the 1/2 for a direct allowed transition. Here $F_R$ has been put into Eq. (2) as $\alpha$.[44] Based on this, the plot of $(\alpha h\nu)^2$ versus $h\nu$ is presented in Fig. 6(b). A representative case of HCO has been extracted from Fig. 6(b) to demonstrate the determination of bandgap, as shown in Fig. 6(c). The region with a linear increase of absorption with increasing energy in Tauc plot is characteristic of semiconductor materials and can be linearly fitted and extrapolated to the x-axis. The estimation of bandgap can thus be obtained by the x-axis intersection point of the linear fit, shown as purple solid line in Fig. 6(c). The bandgaps of all $RCrO_3$ are tabulated in Table 8. It can be noticed that the bandgap for GCO, TbCO, DCO, HCO, ECO, and TmCO is 3.21, 3.27, 3.28, 3.29, 3.28, and 3.29 eV, respectively. These observed values of bandgap are close to some of the reported bandgap of $RCrO_3$.[17] This wide bandgap (~3.3eV) of the present $RCrO_3$ can be possibly attributed to the charge transfer gap of $O^{2-}$ $2p$ -$Cr^{3+}$ $3d(t_{2g})$.[17,45]

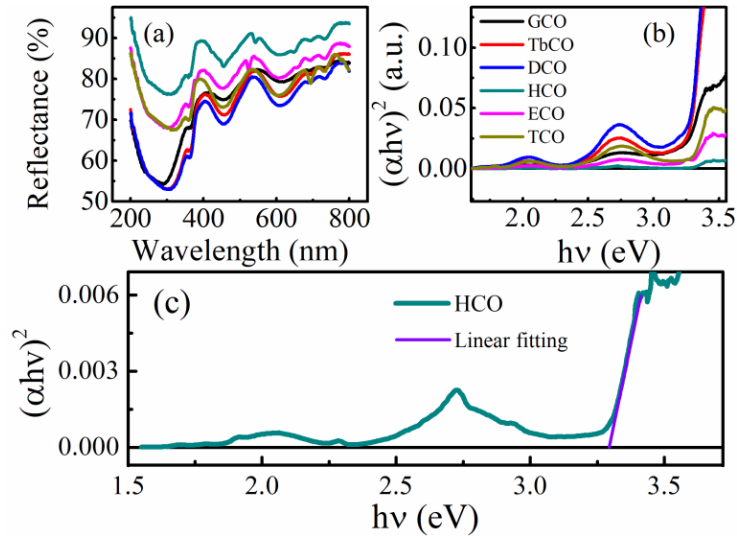

**Figure 6.** (a) UV-Vis diffuse reflectance spectra and (b) optical absorption plots $(\alpha h\nu)^2$ of rare-earth chromites. (c) optical absorption edge of $HoCrO_3$.

The first principles calculations were carried out to understand the electronic and optical properties of $RCrO_3$. We have calculated the bandgaps for the $RCrO_3$ using the spin polarized PBE+U and SCAN+U with VASP. Fig. 7 presents the band structure of $RCrO_3$ calculated by PBE+U. The top of valence band is set to be zero (Fermi level) for convenience. All the members of $RCrO_3$ exhibit very similar band structure. It is observed that $RCrO_3$ is an indirect gap semiconductor as the highest occupied orbitals and the minimum of the lowest unoccupied orbitals both occur at the point S and Γ, respectively, as illustrated in Fig. 7. The density of states (DOS) of the $RCrO_3$ as a function of energy is plotted in Fig. 8(a). As it can be seen in Fig. 8(a), all $RCrO_3$ exhibits similar energy gap with a value around 2.6 eV and listed in Table 8. The obtained value of bandgap for $RCrO_3$ is lower than the experimental one. This is expected as the GGA-PBE always underestimates the electronic bandgap.[46] In the vicinity of Fermi level (- 3 eV – 0 eV) of the valence band of GCO, the majority of the DOS arises from the $d$ states of Cr and $p$ states of O.[30] The bottom of the conduction band is composed of O $2p$ and Cr $3d$ orbitals, but mainly dominated by Cr $3d$ orbitals. Therefore, the observed optical transition originates from O $2p$ orbitals of valence band to Cr $3d$ orbitals of conduction band.

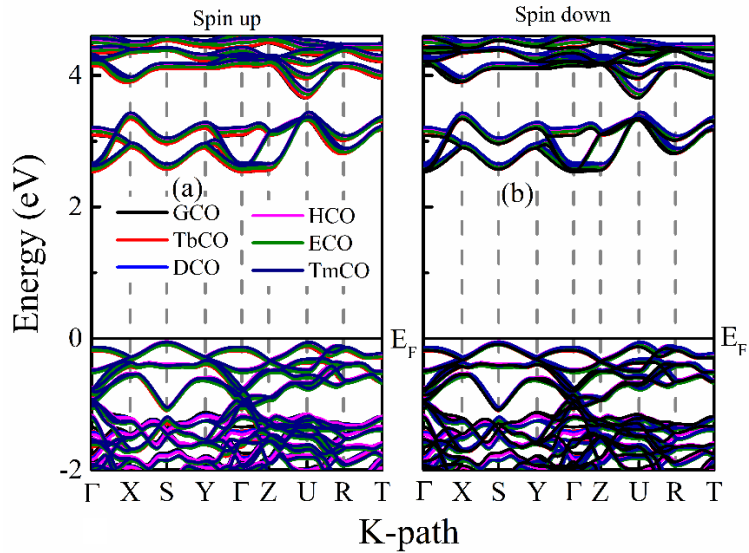

**Figure 7.** Band structure of $RCrO_3$ calculated by PBE-GGA approximation with Hubbard U correction, (a) up-spins and (b) down-spins. The top of valence band is set to be zero (Fermi level, $E_F$).

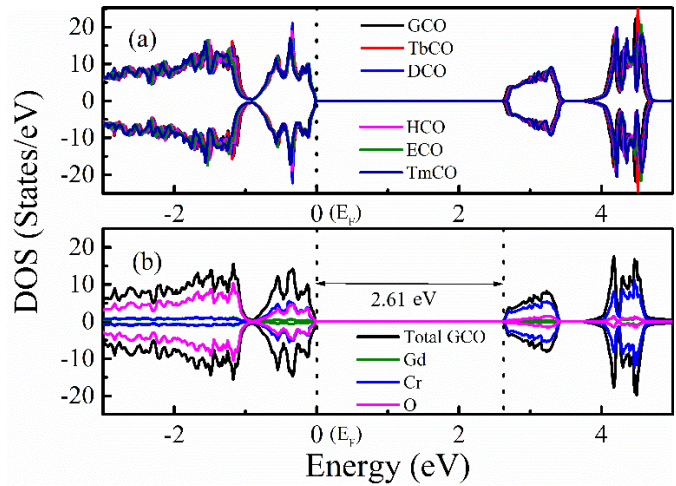

**Figure 8.** (a) Calculated total density of states (DOS) for RCrO$_3$ by the PBE+U simulation. (b) Calculated total and atom projected DOS of GdCrO$_3$ by the PBE+U simulation. The top of valence band is set to be zero (Fermi level, E$_F$).

As observed in Fig. 7, the DFT with PBE+U underestimate the band gap of RCrO$_3$ as compared to that of experimental bad gaps. Therefore, the SCAN has been proposed to calculate the band structure of RCrO$_3$, as SCAN is proved to be better than the PBE version of the GGA exchange correlation functional in reproducing accurate and correct ground-state structures of several other compounds.[47] However, SCAN functional also requires a Hubbard U correction to reproduce the ground-state lattice parameters, magnetic moments, and electronic properties of several materials such as Ce-, Mn-, and Fe-based oxides.[47] To explore whether SCAN+U scheme can give a reasonably accurate description of the electronic structure of the RCrO$_3$, the band structure for RCrO$_3$ were calculated and shown in Fig. 9. The band structure in Fig. 9 also confirms that RCrO$_3$ is an indirect band gap semiconductor, for the valence band maximum (VBM) is at S and the conduction band minimum (CBM) is at Γ.

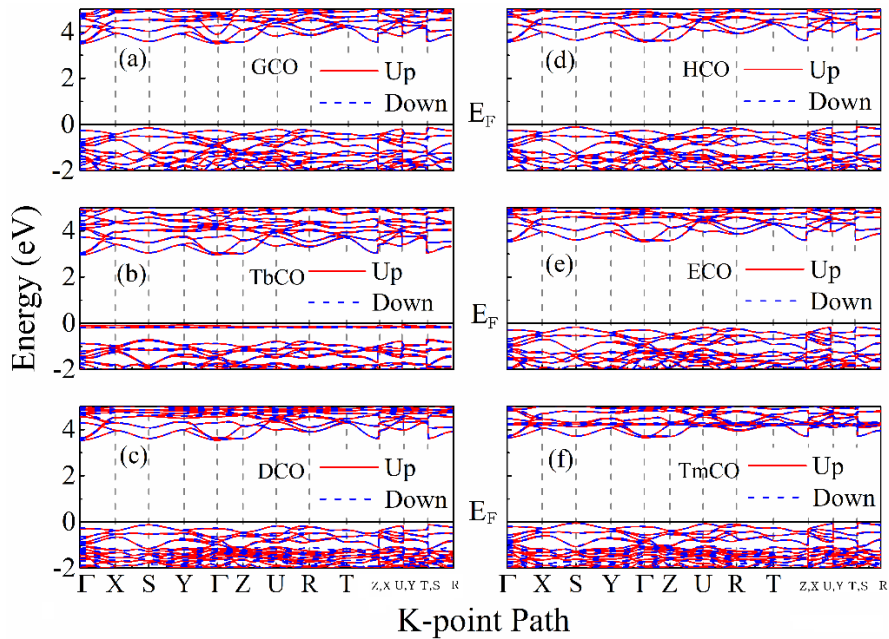

**Figure 9.** Calculated band structure by SCAN with Hubbard U correction, (a) GdCrO$_3$, (b) TbCrO$_3$, (c) DyCrO$_3$, (d) HoCrO$_3$, (e) ErCrO$_3$, (b) TmCrO$_3$. The top of valence band is set to be zero (Fermi level, E$_F$).

In order to elaborate the origin of the electronic band structure in Fig. 9, the computed total and atom projected DOS of RCrO$_3$ are shown in Fig. 10. Fig. 10 shows the density of spin-up and spin-down states of RCrO$_3$ both behave as semiconductor, consistent with previous band structure calculations in Fig. 9. The contribution from R, Cr 3$d$ and O 2$p$ states to the total DOS have been shown and used to determine the type of band gap. The top of the valence band shows an O 2$p$ character, while the bottom of the conduction band has a Cr 3$d$ character.[48] The contribution from O 2$p$ state dominates in the energy range below Fermi level, while the Cr 3$d$ state contribution more above Fermi level in the conduction band. The contribution from R crossing the Fermi level is negligible when compared to the contribution from Cr 3$d$ and O 2$p$ states. According to the charge-transfer (CT) energy required to move an electron from the anion valence band to the $d$ orbitals at the transition-metal site, RCrO$_3$ can be classified as CT semiconductors with a $p - d$ type gap between the O$^{2-}$ 2$p$ filled band and the Cr$^{3+}$ 3$d$ upper Hubbard band.[49] The band gaps of GCO, TbCO, DCO, HCO, ECO, and TmCO are estimated to be 3.46 eV, 2.92 eV, 3.43 eV, 3.53 eV, 3.44 eV, and 3.52 eV from Fig. 10, respectively, as listed in Table 8. Obviously, the present energy gap is wider than the experimentally obtained energy-gap value of ~3.2 eV, and also wider than another calculation result of ~2.7 eV by PBE+U. The calculated results indicate that the SCAN+U could reproduce the Cr 3*d* and O 2*p* bands crossing the Fermi level more correctly than that by PBE+U.

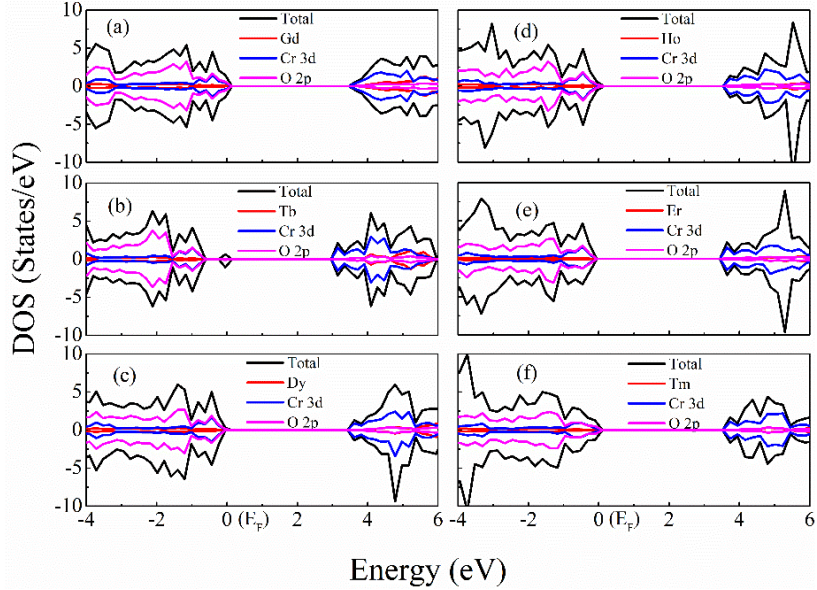

**Figure 10.** The calculated and atom projected density of states by SCAN+U simulation. (a) GdCrO$_3$, (b) TbCrO$_3$, (c) DyCrO$_3$, (d) HoCrO$_3$, (e) ErCrO$_3$, (b) TmCrO$_3$. The top of valence band is set to be zero (Fermi level, E$_F$).

**Table 8.** Bandgap of RCrO$_3$ calculated from the Tauc plot (experimental), and DFT (PBE+U, SCAN+U).

| Sample | Experimental (eV) | DFT (PBE+U) (eV) | DFT (SCAN+U) (eV) |
|---|---|---|---|
| GCO | 3.21 | 2.61 | 3.46 |
| TbCO | 3.27 | 2.62 | 2.92 |
| DCO | 3.28 | 2.63 | 3.43 |
| HCO | 3.29 | 2.65 | 3.53 |
| ECO | 3.28 | 2.66 | 3.44 |
| TmCO | 3.29 | 2.66 | 3.52 |

4. **Conclusion**

A comprehensive study in terms of structural, optical, and electronic properties of RCrO$_3$ (R=Gd, Tb, Dy, Ho, Er, and Tm) has been carried out on all the samples both experimentally and theoretically by first-principles density functional theory. Band gap and structural distortions of RCrO$_3$ in terms of Cr-O-Cr bond angle, unit cell volume, and lattice parameters have been experimentally determined from the UV-Vis spectra and Rietveld refinement, respectively. The lattice parameters *a* and *c* showed a consistent decrease with decreasing ionic radii of rare-earth cation. Raman modes, such as A$_g$(3) and A$_g$(5), shifted to lower numbers with increasing ionic radii of R, indicating a consistent structural distortion associated with octahedral rotation. The optical band gap of RCrO$_3$ revealed by UV-Vis spectra was around 3.3 eV. We also evaluated the performance of generalized gradient approximation (PBE) and meta-GGA (SCAN) functional with Hubbard U correction for predicting the structural and electronic properties in perovskite type $RCrO_3$. The severe crystal distortion with smaller R-site ionic radius was also revealed by both PBE+U and SCAN+U DFT simulations, which agrees with the results observed from XRD and Raman spectra. The SCAN+U framework could reproduce a more accurate ground-state crystal structure than that predicted by PBE+U. The band gap of $RCrO_3$ predicted by SCAN+U simulation (~3.4 eV) is close to the band gap (~3.3 eV) determined experimentally, while the band gap calculated by PBE+U (~ 2.6 eV) is a little bit off from experimental value. SCAN+U has shown to be superior to PBE+U in predicting both the geometries and band structure of $RCrO_3$.

**Acknowledgements:**

Author MJ really acknowledges UConn Institute of Materials Science - Interdisciplinary Multi-Investigator Materials Program (IMS-IMMP) as well as UConn - College of Liberal Arts and Sciences (CLAS) for financial support of this research. The VASP calculations have been performed using the ab-initio total-energy and molecular dynamics program developed at the Fakultät für Physik of the Universität Wien,[32, 33] under a software license agreement between GWF's research group at UConn and the Universität Wien. We also acknowledge the computing resources provided by the Center for Functional Nanomaterials, which is a U.S. DOE Office of Science Facility, at Brookhaven National Laboratory under Contract No. DE-SC0012704. SLS thanks the US Department of Energy, Office of Basic Energy Sciences, Division of Chemical, Biological and Geological Sciences under grant DE-FG02-86ER13622 for support of this research.